\documentclass[prl,twocolumn,groupedaddress,nofootinbib,english,floatfix,superscriptaddress,preprintnumbers]{revtex4-1}
\usepackage{graphicx}
\usepackage{cancel}
\usepackage{amssymb}
\usepackage{textcomp}
\usepackage{amsmath}
\usepackage{mathtools}
\usepackage{bm}
\usepackage{times}
\usepackage{epsfig}
\usepackage[dvipsnames]{xcolor}
\usepackage{graphics}
\usepackage{hyperref}
\usepackage{setspace}
\usepackage{comment}
\usepackage{subcaption}
\usepackage[utf8]{inputenc}
\usepackage[english]{babel}
\usepackage{textcomp}
\usepackage[section]{placeins}
\usepackage{latexsym}
\usepackage{mathrsfs}

\usepackage{tikz}
\usetikzlibrary{patterns}
\usetikzlibrary{shapes.geometric}
\usetikzlibrary{decorations.pathmorphing}
\usetikzlibrary{arrows.meta}
\tikzset{snake it/.style={decorate, decoration=snake}}
\usetikzlibrary{positioning}
\usetikzlibrary{arrows,shapes,positioning}
\usetikzlibrary{decorations.markings}
\tikzstyle arrowstyle=[scale=1]
\tikzstyle directed=[postaction={decorate,decoration={markings,mark=at position .65 with {\arrow[arrowstyle]{stealth}}}}]
\tikzstyle reverse directed=[postaction={decorate,decoration={markings,mark=at position .65 with {\arrowreversed[arrowstyle]{stealth};}}}]

\tikzset{->-/.style={decoration={
  markings,
  mark=at position #1 with {\arrow{>}}},postaction={decorate}}}

\tikzset{-<-/.style={decoration={
  markings,
  mark=at position #1 with {\arrow{<}}},postaction={decorate}}}

\newcommand{\rC}{{\mathrm{C}}}
\newcommand{\rU}{{\mathrm{U}}}
\newcommand{\rI}{{\mathrm{I}}}
\newcommand{\rII}{{\mathrm{II}}}
\newcommand{\rIII}{{\mathrm{III}}}
\newcommand{\rIV}{{\mathrm{IV}}}

\newcommand{\mH}{\ensuremath{\mathcal{H}}}
\newcommand{\mHc}{\ensuremath{\mathcal{H}_c}}
\newcommand{\mCH}{\ensuremath{\mathcal{CH}}}
\newcommand{\ome}{\ensuremath{\omega_{\rI}}}
\newcommand{\omi}{\ensuremath{\omega_{\rII}}}
\newcommand{\ri}{\ensuremath{\mathrm{in}}}
\newcommand{\ru}{\ensuremath{\mathrm{up}}}
\newcommand{\omegap}{\ensuremath{\omega^{+}}}

\newcommand{\VEV}[1]{{\langle #1 \rangle}}
\newcommand{\del}{{\partial}}

\DeclareMathOperator{\csch}{csch}

\begin{document}

\title{Quantum (dis)charge of black hole interiors}

\author{Christiane Klein}
\email{klein@itp.uni-leipzig.de}
\author{Jochen Zahn}
\email{jochen.zahn@itp.uni-leipzig.de}
\author{Stefan Hollands}
\email{stefan.hollands@uni-leipzig.de}
\affiliation{Institut f\"ur Theoretische Physik, Universit\"at Leipzig,\\ Br\"uderstra{\ss}e 16, 04103 Leipzig, Germany}

\begin{abstract}
We analyze the ``vacuum'' polarization induced by a quantum charged scalar field 
near the inner horizon of a charged 
black hole in quantum states evolving from arbitrary regular in states.
Contrary to naive expectations, we find that near an inner horizon, the transversal component of the expected current density  
can have either sign depending on the black hole and field parameters. Thus, the inner horizon can be charged or discharged. But we find that it is always discharged close to extremality.
We also find that quantum effects dominate in that 
the strength of the blow up of the quantum current at the inner horizon is state-independent and stronger than that of the current of a classical solution.  
\end{abstract}
\maketitle

%=================================================================================================%

\section{Introduction}

\begin{figure}
\includegraphics[width=0.45\textwidth]{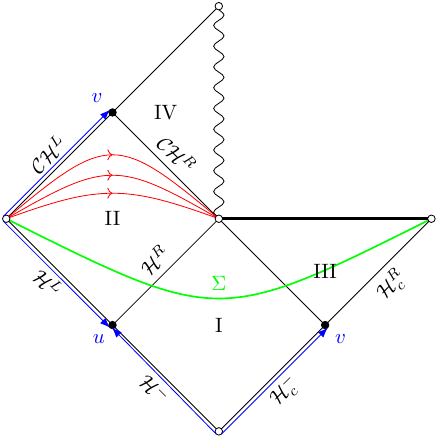}
\caption{\footnotesize Penrose diagram for  Reissner-Nordstr\"om-de Sitter spacetime. $\rI$ and  $\rIII$ constitute the exterior region, with $\rIII$ the region out of causal contact with the BH. Regions $\rII$ and $\rIV$ are the BH interior, which are separated from the exterior by the event horizon $\mH^R$ and from each other by the Cauchy horizon $\mCH^R$. The wiggled line represents the curvature singularity. The green line indicates a Cauchy surface $\Sigma$ for the region $\rI \cup \rII \cup \rIII$, whereas the red lines indicate the direction of the electric field on time-slices in the interior region. The blue arrows indicate the range of the coordinates $u$, $v$ introduced below.
}
\label{fig:RNdS}
\end{figure}

It is well known that  black hole (BH) interiors, such as e.g.\ in the Kerr- or Reissner-Nordstr\" om  
(-deSitter) spacetimes, pose interesting questions regarding determinism. A common feature of these spacetimes is the existence of an ``inner" or ``Cauchy" horizon $\mCH$, beyond which the evolution of classical or quantum fields cannot be predicted from their initial data on a Cauchy surface $\Sigma$ at early times, see \ Fig.~\ref{fig:RNdS}.  

The strong cosmic censorship (sCC) conjecture \cite{Penrose:1974} proposes that the issue is academic because
local observables are expected to diverge at $\mCH$. A more precise formulation of this idea due to \cite{Christodoulou:2008} states that sCC holds if the stress tensor of a classical field, such as a scalar or a linearized gravitational field, is locally not integrable near $\mCH$ for generic smooth initial data on $\Sigma$, thus preventing one from continuing the solution beyond $\mCH$. Hence, one must understand the 
strength of the singularity of physical fields near $\mCH$ to decide whether sCC saves determinism or not.

Estimates for the degree of divergence of an (uncharged) classical Klein-Gordon (KG) field $\Phi$  near the Cauchy horizon of a subextremal Reissner-Nordstr\"om-deSitter (RNdS) spacetime were suggested already in the 1990's \cite{MellorMoss,Mellor:1992,Brady:1998} and have been revisited in recent years \cite{Costa:2014a,Costa:2014b,Costa:2014c, Hintz:2015,Costa:2016,Cardoso:2017}:
\cite{Hintz:2015} have characterized the singular behavior of an arbitrary solution $\Phi$ arising from smooth initial data on $\Sigma$ by its membership in the Sobolev space $H^{\beta + \frac{1}{2}}$ near
$\mCH^R$. Here, $\beta = \frac{\alpha}{\kappa_-}$ where $\alpha$ is the spectral gap of quasinormal modes and $\kappa_-$ is the surface gravity of $\mCH$.
By tracing the dependency of $\beta$ on the BH mass $M$, charge $Q$, and cosmological constant $\Lambda$, it was found 
that $\beta$ can become $> \frac{1}{2}$ \cite{Cardoso:2017}. This implies that the stress tensor of the classical field $\Phi$ is locally  integrable (meaning roughly that $T_{VV} \sim V^{-2+2\beta}$ in terms of a Kruskal coordinate locating $\mCH^R$ at $V=0$, see \ Fig.~\ref{fig:RNdS}), constituting a violation of sCC. 

Given this state of affairs it is of interest to understand whether expectation values of quantum fields diverge near $\mCH^R$ and how the degree of divergence compares to classical fields. 

It was already argued heuristically in the 1970's \cite{BirrellDavies:1978} that the expected stress tensor in a generic quantum state should diverge at $\mCH^R$. This was recently proven \cite{Hollands:2019} for the real KG field in RNdS, in the following sense: in any state $\Psi$ which is regular (i.e. ``Hadamard'') in a neighborhood of a Cauchy surface $\Sigma$ as in \ Fig.~\ref{fig:RNdS}, one has, for $\beta > \frac{1}{2}$,
\begin{equation}
\label{eq:T_VV_Divergence}
   \VEV{T_{VV}}_\Psi \sim C V^{-2},
\end{equation}
with $C$ a constant which only depends on the parameters $M, Q, \Lambda$ of the spacetime and the mass $\mu$ of the field, but not on the quantum state $\Psi$. The state dependence only enters through subleading terms, which are no more singular than the stress tensor for the classical field. The constant $C$ has to be determined numerically and was found to be generically nonzero \cite{Hollands:2020} (see also previous work \cite{Zilberman:2019} on Reissner-Nordstr\" om and special states). Interestingly, both signs of $C$ can occur. 
Assuming that the expected stress tensor backreacts onto the metric via the semiclassical Einstein equation
\begin{equation}
    \label{eq:sEeq}
    G_{\nu\rho}+\Lambda g_{\nu\rho}= 8\pi ( \VEV{ T_{\nu\rho} }_\Psi + E_{\nu \rho} ) ,
\end{equation}
with $E_{\nu \rho}$ the stress tensor of the electromagnetic field, the two signs of $C$ correspond to distinct behavior (infinite stretching resp.\ squeezing) of freely falling observers crossing $\mCH^R$ \cite{Zilberman:2019, Hollands:2019}.
    
As the formation of charged BHs necessitates the presence of charged matter, it is actually more natural to consider a charged scalar field $\Phi$ with the usual minimal coupling $\nabla_\nu \to D_\nu = \nabla_\nu -iqA_\nu$, where $q$ is the charge of the field. It is known \cite{Zhu:2014,Konoplya:2014,Dias:2018, Hod:2018} that for certain values of $(M,Q,\Lambda)$ and $(\mu,q)$, there are classical instabilities (exponentially growing modes) already in the exterior region $\rI$. Excluding such unstable spacetimes, 
we have a positive spectral gap $\alpha>0$ and again $\Phi \in H^{\beta + \frac{1}{2}}$ locally near $\mCH^R$ \cite{Cardoso:2018}. The dependence of $\beta=\alpha/\kappa_-$ on the parameters is influenced by the field charge $q$. There is nevertheless a parameter range--though considerably smaller than for the uncharged scalar field--for which $\beta > \frac{1}{2}$ \cite{Dias:2018, Cardoso:2018}. A proof that \eqref{eq:T_VV_Divergence} also holds for the charged scalar field (with constant $C$ now also $q$ dependent) is presented in Supplemental Material \cite{supplement}, which includes Refs. \cite{Gerard:2014,Verch:1992}.

In the presence of a charged quantum field, backreaction should also take place via the semiclassical Maxwell equation
\begin{equation}
\label{eq:sMeq}
  \nabla^\rho F_{\rho \nu} = -4\pi \VEV{ J_\nu }_\Psi ,
\end{equation}
due to vacuum polarization, i.e., an expectation value for the current $J_\nu$. At the event horizon ${\mH}^R$, this current is responsible for the discharge of the BH via Hawking radiation \cite{Hawking:1974, Gibbons:1975, Unruh:1976}. While vacuum polarization always discharges the event horizon, its influence on the Cauchy horizon is not obvious.
It has been argued invoking the Schwinger process \cite{Schwinger:1951} that pair creation in the interior (region $\rII$) should discharge the Cauchy horizon, i.e., there should be a net current from the left Cauchy horizon $(\mCH^L)$ to the right one $(\mCH^R)$ \cite{Herman:1994,Sorkin:2000}. 

Their arguments are, however, not completely convincing: The BH interior is not stationary, so the very notion of particle is ambiguous there. Furthermore, already at the classical level the behavior of fields at $\mCH^R$ is due to a very nonlocal effect:  
a competition between decay (in region $\rI$) and blue shift (region $\rII$). Hence, Schwinger's formula for pair creation seems not applicable/relevant and a first principle calculation of $\VEV{J_\nu}_\Psi$ at the Cauchy horizon is needed to settle this important question. This is the main novelty presented in our letter. Most interestingly, we find that the relevant component of the current, $\VEV{J_V}_\Psi$, can have either sign at $\mCH^R$, depending on the parameters $Q, M, \Lambda$ of the spacetime and $\mu, q$ of the field. It follows that, via backreaction, the Cauchy horizon can also be charged by quantum effects, contrary to naive expectations \cite{Herman:1994,Sorkin:2000}. 

More precisely, we argue that, in any quantum state $\Psi$ which is initially Hadamard near $\Sigma$, the expectation value of the current diverges at $\mCH^R$, with leading divergence
\begin{equation}
\label{main}
   \VEV{J_V}_\Psi \sim C' V^{-1},    
\end{equation}
where $C'$ is independent of $\Psi$ and can have either sign. The state dependence again enters through subleading terms which behave as the current of a classical field i.e.\ roughly as $J_V \sim V^{-1 + \beta}$.
Hence, as for the stress tensor, quantum effects dominate over classical effects close to the Cauchy horizon. 

Our results thus show that the leading divergence of relevant observables near the Cauchy horizon is of quantum origin and state independent (so that no appeal to ``generic'' initial data is necessary), but also that the ensuing backreaction effects can differ drastically from classical expectations (``mass inflation'') \cite{Poisson:1989}: Not only is a tidal stretching of observers possible (corresponding to ``mass deflation''), but also ``charge inflation''. These possibilities clearly correspond to drastically different forms of the terminal singularity replacing the Cauchy horizon.

\section{Geometric setup and backreaction}
\label{sec:Setup}

The metric $g$ and vector potential $A$ of the RNdS spacetime (Fig.~\ref{fig:RNdS}) are given in natural units $\hbar = c = G_N = 4\pi\epsilon_0=1$ by
\begin{subequations}
\begin{align}
    g&=-f(r) \text{d} t^2+f(r)^{-1} \text{d}r^2 + r^2 \text{d}\Omega^2\\
    f(r)&=-\frac{\Lambda}{3}r^2+1-\frac{2M}{r}+\frac{Q^2}{r^2} \\
    A & = - \frac{Q}{r} \text{d} t,
\end{align}
\end{subequations}
with $\text{d} \Omega^2$ the area element of the unit sphere. The function $f(r)$ has three positive roots $r_c > r_+ > r_-$, corresponding to the cosmological ($\mHc$), event ($\mH$) and Cauchy, horizon ($\mCH$) of the BH.

A useful alternative radial coordinate is the tortoise coordinate $r_*$, defined by $f(r) \text{d}r_* = \text{d}r$. One defines radial null coordinates
   $v \equiv t+r_*$ and $u \equiv t-r_*$.
Their ranges are indicated by blue arrows in Fig.~\ref{fig:RNdS}, pointing from $-\infty$ to $+\infty$.
To extend the metric smoothly across the horizons, 
we use the Kruskal coordinates
$U\equiv \mp e^{-\kappa_+u}$, $V_c\equiv -e^{-\kappa_c v}$, and $ V \equiv -e^{-\kappa_-v}$,
where $\kappa_i=\tfrac{1}{2}| f'(r_i)|$ are the surface gravities on the corresponding horizons. For $U$, we use the $-$ sign in the exterior, and the $+$ sign in the interior, and the $V$ coordinates are defined in the exterior ($V_c$) and interior ($V$) respectively. Note that $V=0$ on $\mCH^R$ and that, by the tensor transformation law, the constants $C,C'$
in \eqref{eq:T_VV_Divergence} and \eqref{main} are given by $C = \kappa_-^{-2} \VEV{T_{vv}}_\Psi$ and $C' = -\kappa_-^{-1} \VEV{J_{v}}_\Psi$, where the expectation values are computed on the inner horizon.
In the following, we present our results in terms of $\VEV{T_{vv}}_\Psi$ and $\VEV{J_{v}}_\Psi$.

Before presenting our arguments in favor of \eqref{main}, let us briefly discuss backreaction effects.
We proceed heuristically and assume that the main contribution to 
backreaction is spherically symmetric, as the leading divergence in \eqref{main}. Thus, for the metric $g$ and the field strength tensor $F$, involving backreaction, we make the ansatz
\begin{align}
   g & = -e^\sigma \text{d}u \text{d}v + r^2 \text{d}\Omega^2, &
   F & = -\frac{Q}{2r^2} e^\sigma \text{d}u \wedge \text{d}v,
\end{align}
where $\sigma$, $r$, and $Q$ are functions of $u$ and $v$. By Gau{\ss}'s law, $Q(u,v)$ is the charge contained within the sphere at $u, v = \text{const}$, which has area $4 \pi r^2$. 

In a first approximation, we now use the expectation value of the current obtained on the RNdS background and obtain, from the $v$ component of \eqref{eq:sMeq},
\begin{equation}
    \del_v Q = - 4 \pi r^2 \VEV{J_v}_\Psi.
\end{equation}
Hence, the sign of $\VEV{ J_v }_\Psi$, evaluated on $\mCH^R$, determines whether quantum effects discharge ($\VEV{J_v}_\Psi > 0$) or charge ($\VEV{J_v}_\Psi < 0$) the BH interior. 
 The resulting behavior of the field strength $\frac{Q}{r^2}$ near $\mCH^R$  is discussed in \cite{supplement}, see also \cite{Zilberman:2019}.

\section{The scalar field and its current}
\label{sec:Scalar}

The charged scalar field $\Phi$ obeys the KG equation
\begin{align}
\label{eq:KG}
    \left[D_\nu D^\nu -\mu^2\right]\Phi&=0, 
\end{align}
The mode ansatz 
\begin{align}
    \Phi_{\ell m} = (4\pi |\omega |)^{-1/2}r^{-1}Y_{\ell m}(\theta ,  \phi ) e^{-i\omega t} H_{\omega\ell}(r)\, ,
\end{align}
with $Y_{\ell m}(\theta,\phi)$ the spherical harmonics, reduces \eqref{eq:KG} to a one-dimensional problem
\begin{subequations}
\begin{align}
\label{eq:Diff_Eq}
    &\left[\partial_{r_*}^2+\left(\omega-\frac{qQ}{r}\right)^2-W\right]H_{\omega \ell}(r_*)=0\\
\label{eq:ScatteringPotential}
    &W=f(r)\left(\frac{\partial_rf(r)}{r}+\frac{\ell (\ell +1)}{r^2}+\mu^2\right)\, .
\end{align}
\end{subequations}
Using gauge transformations of the form
\begin{align}
\label{eq:gauge}
    A &\to A+\frac{Q}{r_0}\, \text{d}t &\Phi&\to e^{i\tfrac{qQ}{r_0}t}\Phi 
\end{align}
we can set the potential to zero at any chosen $r_0$. If $r_0\in\{r_+,r_-,r_c\}$, the solutions $h_{\omega\ell}(r_*,t)=e^{-i\omega t}H_{\omega\ell}(r_*)$ will behave as free waves in that gauge when approaching the corresponding horizon.
We will denote the gauge where $A=0$ at $r_i$, $i\in\{-,+,c\}$, by an $(i)$ superscript .

Following standard procedures \cite{Hollands:2014}, we quantize the field $\Phi$ by choosing a set of positive frequency mode solutions. It is convenient to choose these as corresponding to the Unruh vacuum, which in turn corresponds to certain initial conditions on $\mHc^-\cup\mH^-\cup\mH^L$. There are two types of modes, $h^{(c)\ri}_{\omega\ell}\sim e^{-i\omega V_c}$ on $\mHc^-$,
 and $\sim 0$ on $\mH^-\cup\mH^L$, and $h^{(+)\ru}_{\omega\ell}\sim e^{-i\omega U}$ on $\mH^-\cup\mH^L$ and $\sim 0$ on $\mHc^-$.
The field operator is then given by an expansion in the positive frequency solutions, 
\begin{align}
\Phi(x)&=\int\limits_{0}^{\infty}\text{d}\omega \sum_{\lambda,\ell,m} \left(\Phi_{\omega \ell m}^{\lambda}(x) a_{\omega \ell m }^{\lambda} +\Phi_{-\omega \ell m}^{\lambda}(x)b_{\omega \ell m}^{\lambda \dagger}\right)\, ,
\end{align}
where the coefficients $a_{\omega\ell m}^\lambda$ and $b_{\omega\ell m}^\lambda$ act as annihilation operators on the Unruh vacuum, $a_{\omega\ell m}^\lambda|0\rangle_\rU=b_{\omega\ell m}^\lambda|0\rangle_\rU=0$, and $\lambda$ runs over the types of modes, $\ri$ and $\ru$. That this defines a proper quantum field/state in the absence of classical instabilities, i.e., $\alpha > 0$, can be shown as in \cite{Hollands:2019}, see \cite{supplement}.

The observable driving backreaction onto the electromagnetic field is the current density given by
\begin{align}
    J_\nu(x) = iq\left(\Phi(x)(D_\nu\Phi)^*(x)-\Phi^*(x)D_\nu \Phi(x)\right)\, .
\end{align}
It is local and non-linear in the field, so it requires renormalization in quantum field theory. To evaluate its expectation value in the state $\Psi$, we proceed analogously to recent treatments of the stress-energy tensor: as in \cite{Hollands:2019}, 
the expectation value in the Unruh state $\VEV{J_V}_\rU$ is already found 
to give the leading contribution to \eqref{main}, because the $\Psi$-dependent contributions are subleading. In \cite{supplement}, this is shown rigorously for $\beta > \frac{1}{2}$, analogously to \cite{Hollands:2019}, but is expected to hold also for $\beta < \frac{1}{2}$ in a large class of states. 

To evaluate $\VEV{J_v}_\rU$ we use a Hadamard point-split renormalization, similar to the one performed in \cite{Lanir:2017,Sela:2018,Zilberman:2019,Hollands:2019} for the stress tensor. By a suitable point-split prescription, all singularities of the coinciding point limit vanish, so that Hadamard point-split renormalization amounts to the subtraction of a finite part, which actually vanishes on the horizons \cite{PRDpaper}. We then calculate the $v$ component of the current on $\mathcal{CH}^L$, which coincides with the value on $\mathcal{CH}^R$ because the state is stationary, by taking the limit of \cite[eq. (39)]{PRDpaper} onto the horizon $\mCH^L$.

The "Boulware modes" defined in \cite{PRDpaper} behave as
\begin{subequations}
\label{eq:asymp}
\begin{align}
\tilde h^{(+) \ri , \rI}_{\omega \ell }&=
\mathcal{T}^\rI_{(\omega-\ome)\ell}\frac{\omega}{\omega-\ome}e^{-i(\omega-\omega_\rI) v} \quad \text{ on } \mH^R  \\
\tilde h^{(+) \ru,\rI}_{\omega \ell m}&=
\mathcal{R}^\rI_{\omega\ell} e^{-i\omega v} \qquad \qquad \qquad \text{ on }\mH^R\\
\tilde h^{(-) \ri,\rII}_{\omega \ell m}&=
\overline{\mathcal{T}^{\rII}_{\omega\ell}}e^{-i(\omega-\omi) v} \qquad \qquad \text{ on }\mCH^L\\ 
\tilde h^{(-) \ru,\rII}_{\omega \ell m}&=\mathcal{R}^{\rII}_{\omega\ell} e^{-i(\omega-\omi) v} \qquad \qquad \text{ on }\mCH^L\,,
\end{align}
\end{subequations}
where $\ome=qQ(r_+^{-1}-r_c^{-1})$ and $\omi=qQ(r_-^{-1}-r_+^{-1})$. This follows from  the differential equation (\ref{eq:Diff_Eq}), and the relation of the modes. Here, $\mathcal{T}^{\mathrm{N}}_{\omega\ell}$ and $\mathcal{R}^{\mathrm{N}}_{\omega\ell}$ are the transmission and reflection coefficients of the $\tilde{h}^{\ru,\mathrm{N}}_{\omega\ell}$ modes \cite[eq. (16), (17)]{PRDpaper}. The scattering coefficients of the $\ri$-type modes have been expressed in terms of these by comparing the behaviour of the modes for $r_*\to \pm\infty$. 

Inserting this into \cite[eq. (39)]{PRDpaper}, one finds
\begin{subequations}
\begin{align}
\label{eq:current_fin_sum}
 \VEV{J_v}_\rU & = -\sum\limits_{\ell=0}^{\infty}\frac{q(2\ell+1)}{16\pi^2r_-^2}\int\limits_{0}^{\infty}\text{d}\omega\left[F_\ell(\omega)+F_\ell(-\omega)\right], \\ \nonumber
 F_\ell(\omega) & =\frac{\omega(\omegap+\ome)}{(\omegap)^2}\coth\left(\pi\tfrac{\omegap+\ome}{\kappa_c}\right) \left|\mathcal{T}^\rI_{\omegap\ell}\right|^2\left|\mathcal{T}^{\rII}_{\omegap\ell}\right|^2\\\nonumber
&+\frac{\omega\coth\left(\pi\tfrac{\omegap}{\kappa_+}\right)}{\omegap}\left[\left|\mathcal{R}^{\rII}_{\omegap\ell}\right|^2 + \left|\mathcal{R}^\rI_{\omegap\ell}\right|^2 \left|\mathcal{T}^{\rII}_{\omegap\ell}\right|^2\right]\\
\label{eq:integrand}
&+\frac{2\omega \csch \left(\pi\tfrac{\omegap}{\kappa_+}\right)}{\omegap} \mathrm{Re}\left(\overline{\mathcal{R}^\rI_{\omegap\ell}} \mathcal{T}^{\rII}_{\omegap\ell} \mathcal{R}^{\rII}_{\omegap\ell}\right),
\end{align}
\end{subequations}
with $\omegap=\omega+\omi$.
A similar result for $\VEV{T_{vv}}_\rU$ is given in \cite{supplement}.

%========================================================================================%
\section{Numerical results}
\label{sec:NR}
To compute the scattering coefficients, \eqref{eq:Diff_Eq} has to be solved numerically. Following \cite{Suzuki:1999}, in the case of a conformal mass $\mu^2 = \frac{2 \Lambda}{3}$, one can bring the equation into the form of a Heun differential equation by a special choice of ansatz. The solutions to the Heun equation have been implemented in {\it Mathematica} as special functions \cite{Hatsuda:2020}. For a general $\mu^2$, the same ansatz yields a slightly more general equation, that can however be solved, near any of the horizons, by a power series as described in \cite{Hollands:2020}. By comparing the solutions around different horizons, one can calculate the scattering coefficients \cite{Hollands:2020}.

For small enough $\mu$, the integrand in \eqref{eq:current_fin_sum} is rapidly decreasing both in $\omega$ and $\ell$. We thus calculate the scattering coefficients and the corresponding $F_\ell(\omega)$, c.f.\ \eqref{eq:integrand}, for different values of $\omega$ and $\ell$ up to some maximal values beyond which the contribution to \eqref{eq:current_fin_sum} is negligible. 
Then, the integral in \eqref{eq:current_fin_sum} is estimated by a Riemann sum, which is also used to estimate the errors of this procedure.

\begin{figure}[h]
    \centering
    \includegraphics[width=0.45\textwidth]{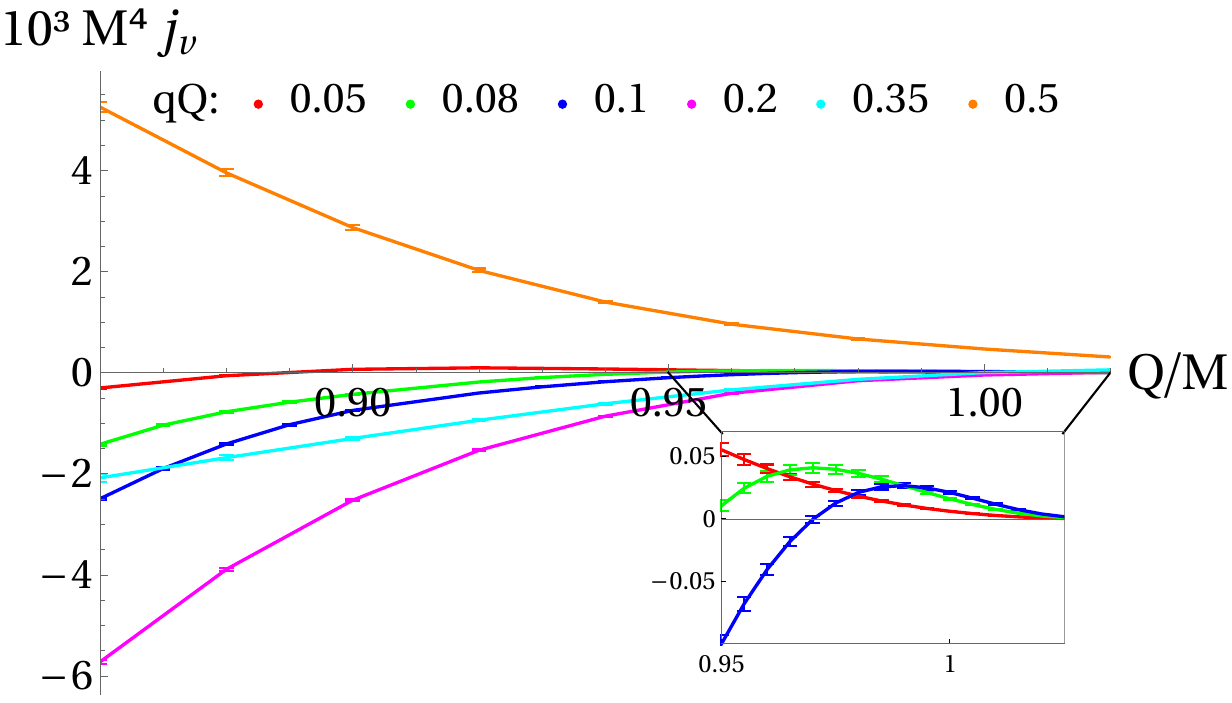}
    \caption{\footnotesize $\VEV{J_v}_\rU$ evaluated on $\mCH^R$ as a function of $Q/M$ for different values of $qQ$ and $\mu^2=2\Lambda/3$, $\Lambda=0.14 M^{-2}$.}
    \label{fig:QscanW}
\end{figure}

Figure~\ref{fig:QscanW} shows the results for the current component $J_v$ on $\mCH^R$ as a function of $Q/M$ (nearly up to extremality) for different charges of the scalar field. The most remarkable feature is that the current can have either sign, depending on the charge of the BH and of the field. As discussed above, $\VEV{J_v}_\rU > 0$ corresponds to the discharge of the Cauchy horizon, while for $\VEV{J_v}_\rU < 0$ backreaction effects increase its charge. We see that, close enough to extremality, backreaction effects discharge the Cauchy horizon, so that it is driven away from extremality. However, away from extremality, a sign change may occur, so that quantum effects can tend to charge Cauchy horizons of BHs which are far enough away from extremality. One may also check that, for charges $q$ small enough and all other parameters fixed, the current is proportional to $q^2$, i.e. the fine-structure constant, compatible with expectations.

The possibility of a sign change of the current at the inner horizon is in stark contrast to the situation at the outer horizon, where vacuum polarization always tends to discharge the BH \cite{PRDpaper,Gibbons:1975} in physically reasonable states. We attribute the phenomenon to the scattering of modes entering the BH through the event horizon off the potential barrier \eqref{eq:ScatteringPotential}. Note, however, that it is impossible to clearly distinguish the contributions from the inflow of a current through the event horizon and the creation of a current in the BH interior, as the up modes in the interior, which are responsible for the latter effect, are entangled with the up modes in the exterior region, leading to interference effects, manifest in the last term in \eqref{eq:integrand}.

\begin{figure}[h]
   \centering
   \includegraphics[width=0.45\textwidth]{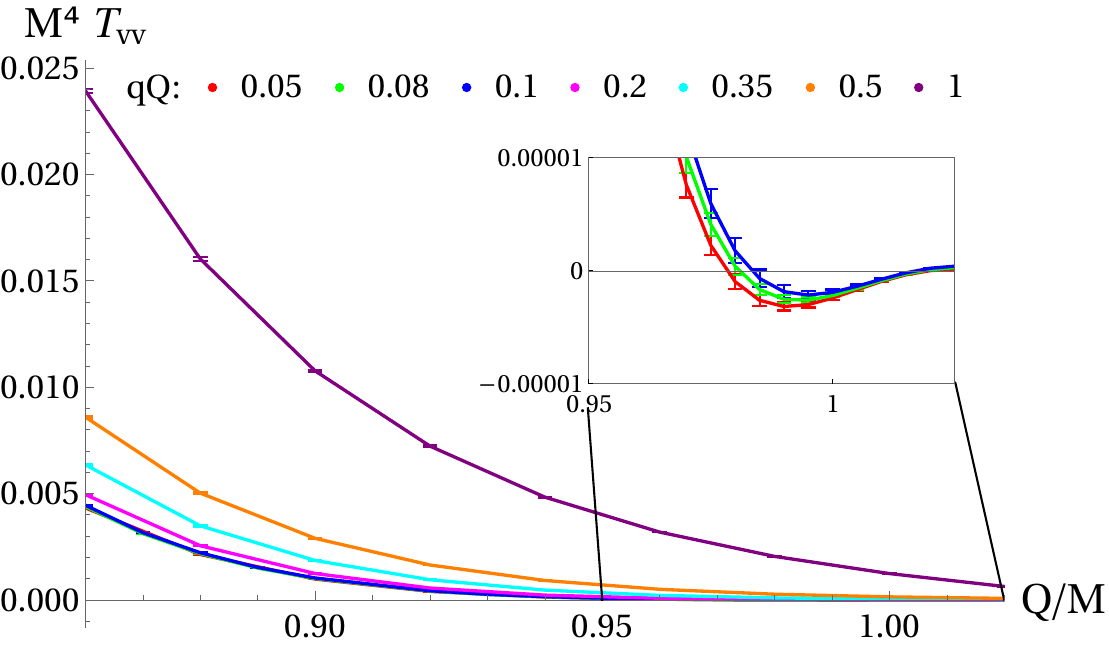}
   \caption{\footnotesize $\VEV{T_{vv}}_{\rU-\rC}$ evaluated on $\mCH^R$ as a function of $Q/M$ for different values of $qQ$ and $\mu^2=2\Lambda/3$, $\Lambda=0.14 M^{-2}$.
   }
    \label{fig:TscanW}
\end{figure}

Evaluation of the stress tensor at $\mCH^R$ \cite{supplement} yields results compatible with those obtained for the real scalar field in \cite{Zilberman:2019, Hollands:2020}, see Fig. \ref{fig:TscanW} and \cite{supplement} for a detailed discussion.
Combining the results for the current and the stress tensor, we find that, for the parameters considered in Fig.~\ref{fig:QscanW} and in the weak backreaction regime, where $\partial_v r=-4\pi r_-/\kappa_-\langle T_{vv}\rangle_\Psi$ \cite{supplement, Zilberman:2019}, backreaction effects can increase the field strength, $\del_v (Q/r^2) > 0$, as one approaches $\mCH^R$, even for parameters for which backreaction discharges the inner horizon \cite{supplement}.

We note that the parameter range considered in Fig.~\ref{fig:QscanW} does not capture semirealistic BH and field parameters. 
To achieve comparability with \cite{Hollands:2020, Cardoso:2017}, a reasonable performance of the numerical code, and to avoid the classical instability
\cite{Dias:2018}, $\Lambda$ was chosen unrealistically high, so that the event horizon $r_+$ of the BH is of the same order of magnitude as the cosmological horizon $r_c$. For our fixed value of $\Lambda$, the bound $Q/M > 0.755$ is necessary to achieve $r_+ < r_c$. But astrophysical BHs are expected to be only weakly charged \cite{Gibbons:1975}, i.e., $Q \ll M$. Furthermore, if we identify $q$ with the elementary charge $e$, we find that for near-extremal BHs $qQ\sim 10^{36} M/M_{\odot}$,
which is beyond the regime accessible to our numerical code for realistic BH masses. Similarly, the assumption of a conformal mass for $\Phi$ is unrealistic, but semirealistic masses $\mu$ are not accessible by our code (however, we checked stability of our results under deviations from the conformal mass). Hence, our results unfortunately cannot be used to infer the fate of the Cauchy horizon of semirealistic BHs. However, they are sufficient to demonstrate that the naive expectation that quantum effects will always discharge the BH interior is false. It would be interesting to see whether a similar effect occurs for rotating (Kerr) black holes where (dis-)charging
would now correspond to (down-)upspinning of the Cauchy horizon and $J_v$ would correspond to $T_{v\phi}$.

%=========================================================================================%
\begin{acknowledgments}
{\bf Acknowledgements:} SH thanks Ted Jacobson for stimulating discussions at an early stage of this project. This work has been funded by the Deutsche Forschungsgemeinschaft (DFG) under the Grant No. 406116891 within the Research Training Group RTG 2522/1.
\end{acknowledgments}

\bibliography{cs}

\begin{thebibliography}{36}%
\makeatletter
\providecommand \@ifxundefined [1]{%
 \@ifx{#1\undefined}
}%
\providecommand \@ifnum [1]{%
 \ifnum #1\expandafter \@firstoftwo
 \else \expandafter \@secondoftwo
 \fi
}%
\providecommand \@ifx [1]{%
 \ifx #1\expandafter \@firstoftwo
 \else \expandafter \@secondoftwo
 \fi
}%
\providecommand \natexlab [1]{#1}%
\providecommand \enquote  [1]{``#1''}%
\providecommand \bibnamefont  [1]{#1}%
\providecommand \bibfnamefont [1]{#1}%
\providecommand \citenamefont [1]{#1}%
\providecommand \href@noop [0]{\@secondoftwo}%
\providecommand \href [0]{\begingroup \@sanitize@url \@href}%
\providecommand \@href[1]{\@@startlink{#1}\@@href}%
\providecommand \@@href[1]{\endgroup#1\@@endlink}%
\providecommand \@sanitize@url [0]{\catcode `\\12\catcode `\$12\catcode
  `\&12\catcode `\#12\catcode `\^12\catcode `\_12\catcode `\%12\relax}%
\providecommand \@@startlink[1]{}%
\providecommand \@@endlink[0]{}%
\providecommand \url  [0]{\begingroup\@sanitize@url \@url }%
\providecommand \@url [1]{\endgroup\@href {#1}{\urlprefix }}%
\providecommand \urlprefix  [0]{URL }%
\providecommand \Eprint [0]{\href }%
\providecommand \doibase [0]{http://dx.doi.org/}%
\providecommand \selectlanguage [0]{\@gobble}%
\providecommand \bibinfo  [0]{\@secondoftwo}%
\providecommand \bibfield  [0]{\@secondoftwo}%
\providecommand \translation [1]{[#1]}%
\providecommand \BibitemOpen [0]{}%
\providecommand \bibitemStop [0]{}%
\providecommand \bibitemNoStop [0]{.\EOS\space}%
\providecommand \EOS [0]{\spacefactor3000\relax}%
\providecommand \BibitemShut  [1]{\csname bibitem#1\endcsname}%
\let\auto@bib@innerbib\@empty
%</preamble>
\bibitem [{\citenamefont {Penrose}(1974)}]{Penrose:1974}%
  \BibitemOpen
  \bibfield  {author} {\bibinfo {author} {\bibfnamefont {R.}~\bibnamefont
  {Penrose}},\ }\enquote {\bibinfo {title} {Gravitational radiation and
  gravitational collapse},}\ \ (\bibinfo  {publisher} {Springer},\ \bibinfo
  {address} {Heidelberg},\ \bibinfo {year} {1974})\ Chap.\ \bibinfo {chapter}
  {Gravitational collapse}\BibitemShut {NoStop}%
\bibitem [{\citenamefont {Christodoulou}(2009)}]{Christodoulou:2008}%
  \BibitemOpen
  \bibfield  {author} {\bibinfo {author} {\bibfnamefont {D.}~\bibnamefont
  {Christodoulou}},\ }\href@noop {} {\emph {\bibinfo {title} {{The Formation of
  Black Holes in General Relativity}}}}\ (\bibinfo  {publisher} {European
  Mathematical Society Publishing House},\ \bibinfo {address} {Zürich},\
  \bibinfo {year} {2009})\ \Eprint {http://arxiv.org/abs/0805.3880}
  {arXiv:0805.3880 [gr-qc]} \BibitemShut {NoStop}%
\bibitem [{\citenamefont {Mellor}\ and\ \citenamefont
  {Moss}(1990)}]{MellorMoss}%
  \BibitemOpen
  \bibfield  {author} {\bibinfo {author} {\bibfnamefont {F.}~\bibnamefont
  {Mellor}}\ and\ \bibinfo {author} {\bibfnamefont {I.}~\bibnamefont {Moss}},\
  }\href {\doibase 10.1103/PhysRevD.41.403} {\bibfield  {journal} {\bibinfo
  {journal} {Phys. Rev. D}\ }\textbf {\bibinfo {volume} {41}},\ \bibinfo
  {pages} {403} (\bibinfo {year} {1990})}\BibitemShut {NoStop}%
\bibitem [{\citenamefont {Mellor}\ and\ \citenamefont
  {Moss}(1992)}]{Mellor:1992}%
  \BibitemOpen
  \bibfield  {author} {\bibinfo {author} {\bibfnamefont {F.}~\bibnamefont
  {Mellor}}\ and\ \bibinfo {author} {\bibfnamefont {I.}~\bibnamefont {Moss}},\
  }\href {\doibase 10.1088/0264-9381/9/4/001} {\bibfield  {journal} {\bibinfo
  {journal} {Classical and Quantum Gravity}\ }\textbf {\bibinfo {volume} {9}},\
  \bibinfo {pages} {L43} (\bibinfo {year} {1992})}\BibitemShut {NoStop}%
\bibitem [{\citenamefont {Brady}\ \emph {et~al.}(1998)\citenamefont {Brady},
  \citenamefont {Moss},\ and\ \citenamefont {Myers}}]{Brady:1998}%
  \BibitemOpen
  \bibfield  {author} {\bibinfo {author} {\bibfnamefont {P.~R.}\ \bibnamefont
  {Brady}}, \bibinfo {author} {\bibfnamefont {I.~G.}\ \bibnamefont {Moss}}, \
  and\ \bibinfo {author} {\bibfnamefont {R.~C.}\ \bibnamefont {Myers}},\ }\href
  {\doibase 10.1103/PhysRevLett.80.3432} {\bibfield  {journal} {\bibinfo
  {journal} {Phys.\ Rev.\ Lett.}\ }\textbf {\bibinfo {volume} {80}},\ \bibinfo
  {pages} {3432} (\bibinfo {year} {1998})},\ \Eprint
  {http://arxiv.org/abs/gr-qc/9801032} {arXiv:gr-qc/9801032} \BibitemShut
  {NoStop}%
\bibitem [{\citenamefont {Costa}\ \emph
  {et~al.}(2015{\natexlab{a}})\citenamefont {Costa}, \citenamefont {Girão},
  \citenamefont {Natário},\ and\ \citenamefont {Silva}}]{Costa:2014a}%
  \BibitemOpen
  \bibfield  {author} {\bibinfo {author} {\bibfnamefont {J.~L.}\ \bibnamefont
  {Costa}}, \bibinfo {author} {\bibfnamefont {P.~M.}\ \bibnamefont {Girão}},
  \bibinfo {author} {\bibfnamefont {J.}~\bibnamefont {Natário}}, \ and\
  \bibinfo {author} {\bibfnamefont {J.~D.}\ \bibnamefont {Silva}},\ }\href
  {\doibase 10.1088/0264-9381/32/1/015017} {\bibfield  {journal} {\bibinfo
  {journal} {Class. Quant. Grav.}\ }\textbf {\bibinfo {volume} {32}},\ \bibinfo
  {pages} {015017} (\bibinfo {year} {2015}{\natexlab{a}})},\ \Eprint
  {http://arxiv.org/abs/1406.7245} {arXiv:1406.7245 [gr-qc]} \BibitemShut
  {NoStop}%
\bibitem [{\citenamefont {Costa}\ \emph
  {et~al.}(2015{\natexlab{b}})\citenamefont {Costa}, \citenamefont {Girão},
  \citenamefont {Natário},\ and\ \citenamefont {Silva}}]{Costa:2014b}%
  \BibitemOpen
  \bibfield  {author} {\bibinfo {author} {\bibfnamefont {J.~L.}\ \bibnamefont
  {Costa}}, \bibinfo {author} {\bibfnamefont {P.~M.}\ \bibnamefont {Girão}},
  \bibinfo {author} {\bibfnamefont {J.}~\bibnamefont {Natário}}, \ and\
  \bibinfo {author} {\bibfnamefont {J.~D.}\ \bibnamefont {Silva}},\ }\href
  {\doibase 10.1007/s00220-015-2433-6} {\bibfield  {journal} {\bibinfo
  {journal} {Commun. Math. Phys.}\ }\textbf {\bibinfo {volume} {339}},\
  \bibinfo {pages} {903} (\bibinfo {year} {2015}{\natexlab{b}})},\ \Eprint
  {http://arxiv.org/abs/1406.7253} {arXiv:1406.7253 [gr-qc]} \BibitemShut
  {NoStop}%
\bibitem [{\citenamefont {Costa}\ \emph {et~al.}(2017)\citenamefont {Costa},
  \citenamefont {Girão}, \citenamefont {Natário},\ and\ \citenamefont
  {Silva}}]{Costa:2014c}%
  \BibitemOpen
  \bibfield  {author} {\bibinfo {author} {\bibfnamefont {J.~L.}\ \bibnamefont
  {Costa}}, \bibinfo {author} {\bibfnamefont {P.~M.}\ \bibnamefont {Girão}},
  \bibinfo {author} {\bibfnamefont {J.}~\bibnamefont {Natário}}, \ and\
  \bibinfo {author} {\bibfnamefont {J.~D.}\ \bibnamefont {Silva}},\ }\href
  {\doibase 10.1007/s40818-017-0028-6} {\bibfield  {journal} {\bibinfo
  {journal} {Annals of PDE}\ }\textbf {\bibinfo {volume} {3}} (\bibinfo {year}
  {2017}),\ 10.1007/s40818-017-0028-6},\ \Eprint
  {http://arxiv.org/abs/1406.7261} {arXiv:1406.7261 [gr-qc]} \BibitemShut
  {NoStop}%
\bibitem [{\citenamefont {Hintz}\ and\ \citenamefont
  {Vasy}(2017)}]{Hintz:2015}%
  \BibitemOpen
  \bibfield  {author} {\bibinfo {author} {\bibfnamefont {P.}~\bibnamefont
  {Hintz}}\ and\ \bibinfo {author} {\bibfnamefont {A.}~\bibnamefont {Vasy}},\
  }\href {\doibase 10.1063/1.4996575} {\bibfield  {journal} {\bibinfo
  {journal} {J.\ Math.\ Phys.}\ }\textbf {\bibinfo {volume} {58}},\ \bibinfo
  {pages} {081509} (\bibinfo {year} {2017})},\ \Eprint
  {http://arxiv.org/abs/1512.08004} {arXiv:1512.08004 [math.AP]} \BibitemShut
  {NoStop}%
\bibitem [{\citenamefont {Costa}\ and\ \citenamefont
  {Franzen}(2017)}]{Costa:2016}%
  \BibitemOpen
  \bibfield  {author} {\bibinfo {author} {\bibfnamefont {J.~L.}\ \bibnamefont
  {Costa}}\ and\ \bibinfo {author} {\bibfnamefont {A.~T.}\ \bibnamefont
  {Franzen}},\ }\href {\doibase 10.1007/s00023-017-0592-z} {\bibfield
  {journal} {\bibinfo  {journal} {Annales Henri Poincare}\ }\textbf {\bibinfo
  {volume} {18}},\ \bibinfo {pages} {3371} (\bibinfo {year} {2017})},\ \Eprint
  {http://arxiv.org/abs/1607.01018} {arXiv:1607.01018 [gr-qc]} \BibitemShut
  {NoStop}%
\bibitem [{\citenamefont {Cardoso}\ \emph
  {et~al.}(2018{\natexlab{a}})\citenamefont {Cardoso}, \citenamefont {Costa},
  \citenamefont {Destounis}, \citenamefont {Hintz},\ and\ \citenamefont
  {Jansen}}]{Cardoso:2017}%
  \BibitemOpen
  \bibfield  {author} {\bibinfo {author} {\bibfnamefont {V.}~\bibnamefont
  {Cardoso}}, \bibinfo {author} {\bibfnamefont {J.~a.~L.}\ \bibnamefont
  {Costa}}, \bibinfo {author} {\bibfnamefont {K.}~\bibnamefont {Destounis}},
  \bibinfo {author} {\bibfnamefont {P.}~\bibnamefont {Hintz}}, \ and\ \bibinfo
  {author} {\bibfnamefont {A.}~\bibnamefont {Jansen}},\ }\href {\doibase
  10.1103/PhysRevLett.120.031103} {\bibfield  {journal} {\bibinfo  {journal}
  {Phys. Rev. Lett.}\ }\textbf {\bibinfo {volume} {120}},\ \bibinfo {pages}
  {031103} (\bibinfo {year} {2018}{\natexlab{a}})},\ \Eprint
  {http://arxiv.org/abs/1711.10502} {arXiv:1711.10502 [gr-qc]} \BibitemShut
  {NoStop}%
\bibitem [{\citenamefont {Birrell}\ and\ \citenamefont
  {Davies}(1978)}]{BirrellDavies:1978}%
  \BibitemOpen
  \bibfield  {author} {\bibinfo {author} {\bibfnamefont {N.}~\bibnamefont
  {Birrell}}\ and\ \bibinfo {author} {\bibfnamefont {P.}~\bibnamefont
  {Davies}},\ }\href {\doibase 10.1038/272035a0} {\bibfield  {journal}
  {\bibinfo  {journal} {Nature}\ }\textbf {\bibinfo {volume} {272}},\ \bibinfo
  {pages} {35} (\bibinfo {year} {1978})}\BibitemShut {NoStop}%
\bibitem [{\citenamefont {Hollands}\ \emph
  {et~al.}(2020{\natexlab{a}})\citenamefont {Hollands}, \citenamefont {Wald},\
  and\ \citenamefont {Zahn}}]{Hollands:2019}%
  \BibitemOpen
  \bibfield  {author} {\bibinfo {author} {\bibfnamefont {S.}~\bibnamefont
  {Hollands}}, \bibinfo {author} {\bibfnamefont {R.~M.}\ \bibnamefont {Wald}},
  \ and\ \bibinfo {author} {\bibfnamefont {J.}~\bibnamefont {Zahn}},\ }\href
  {\doibase 10.1088/1361-6382/ab8052} {\bibfield  {journal} {\bibinfo
  {journal} {Class. Quant. Grav.}\ }\textbf {\bibinfo {volume} {37}},\ \bibinfo
  {pages} {115009} (\bibinfo {year} {2020}{\natexlab{a}})},\ \Eprint
  {http://arxiv.org/abs/1912.06047} {arXiv:1912.06047 [gr-qc]} \BibitemShut
  {NoStop}%
\bibitem [{\citenamefont {Hollands}\ \emph
  {et~al.}(2020{\natexlab{b}})\citenamefont {Hollands}, \citenamefont {Klein},\
  and\ \citenamefont {Zahn}}]{Hollands:2020}%
  \BibitemOpen
  \bibfield  {author} {\bibinfo {author} {\bibfnamefont {S.}~\bibnamefont
  {Hollands}}, \bibinfo {author} {\bibfnamefont {C.}~\bibnamefont {Klein}}, \
  and\ \bibinfo {author} {\bibfnamefont {J.}~\bibnamefont {Zahn}},\ }\href
  {\doibase 10.1103/PhysRevD.102.085004} {\bibfield  {journal} {\bibinfo
  {journal} {Phys. Rev. D}\ }\textbf {\bibinfo {volume} {102}},\ \bibinfo
  {pages} {085004} (\bibinfo {year} {2020}{\natexlab{b}})},\ \Eprint
  {http://arxiv.org/abs/2006.10991} {arXiv:2006.10991 [gr-qc]} \BibitemShut
  {NoStop}%
\bibitem [{\citenamefont {Zilberman}\ \emph {et~al.}(2020)\citenamefont
  {Zilberman}, \citenamefont {Levi},\ and\ \citenamefont
  {Ori}}]{Zilberman:2019}%
  \BibitemOpen
  \bibfield  {author} {\bibinfo {author} {\bibfnamefont {N.}~\bibnamefont
  {Zilberman}}, \bibinfo {author} {\bibfnamefont {A.}~\bibnamefont {Levi}}, \
  and\ \bibinfo {author} {\bibfnamefont {A.}~\bibnamefont {Ori}},\ }\href
  {\doibase 10.1103/PhysRevLett.124.171302} {\bibfield  {journal} {\bibinfo
  {journal} {Phys. Rev. Lett.}\ }\textbf {\bibinfo {volume} {124}},\ \bibinfo
  {pages} {171302} (\bibinfo {year} {2020})},\ \Eprint
  {http://arxiv.org/abs/1906.11303} {arXiv:1906.11303 [gr-qc]} \BibitemShut
  {NoStop}%
\bibitem [{\citenamefont {Zhu}\ \emph {et~al.}(2014)\citenamefont {Zhu},
  \citenamefont {Zhang}, \citenamefont {Pellicer}, \citenamefont {Wang},\ and\
  \citenamefont {Abdalla}}]{Zhu:2014}%
  \BibitemOpen
  \bibfield  {author} {\bibinfo {author} {\bibfnamefont {Z.}~\bibnamefont
  {Zhu}}, \bibinfo {author} {\bibfnamefont {S.-J.}\ \bibnamefont {Zhang}},
  \bibinfo {author} {\bibfnamefont {C.~E.}\ \bibnamefont {Pellicer}}, \bibinfo
  {author} {\bibfnamefont {B.}~\bibnamefont {Wang}}, \ and\ \bibinfo {author}
  {\bibfnamefont {E.}~\bibnamefont {Abdalla}},\ }\href {\doibase
  10.1103/PhysRevD.90.044042} {\bibfield  {journal} {\bibinfo  {journal} {Phys.
  Rev. D}\ }\textbf {\bibinfo {volume} {90}},\ \bibinfo {pages} {044042}
  (\bibinfo {year} {2014})},\ \bibinfo {note} {[Addendum: Phys.Rev.D 90, 049904
  (2014)]},\ \Eprint {http://arxiv.org/abs/1405.4931} {arXiv:1405.4931
  [hep-th]} \BibitemShut {NoStop}%
\bibitem [{\citenamefont {Konoplya}\ and\ \citenamefont
  {Zhidenko}(2014)}]{Konoplya:2014}%
  \BibitemOpen
  \bibfield  {author} {\bibinfo {author} {\bibfnamefont {R.~A.}\ \bibnamefont
  {Konoplya}}\ and\ \bibinfo {author} {\bibfnamefont {A.}~\bibnamefont
  {Zhidenko}},\ }\href {\doibase 10.1103/PhysRevD.90.064048} {\bibfield
  {journal} {\bibinfo  {journal} {Phys. Rev. D}\ }\textbf {\bibinfo {volume}
  {90}},\ \bibinfo {pages} {064048} (\bibinfo {year} {2014})},\ \Eprint
  {http://arxiv.org/abs/1406.0019} {arXiv:1406.0019 [hep-th]} \BibitemShut
  {NoStop}%
\bibitem [{\citenamefont {Dias}\ \emph {et~al.}(2019)\citenamefont {Dias},
  \citenamefont {Reall},\ and\ \citenamefont {Santos}}]{Dias:2018}%
  \BibitemOpen
  \bibfield  {author} {\bibinfo {author} {\bibfnamefont {O.~J.}\ \bibnamefont
  {Dias}}, \bibinfo {author} {\bibfnamefont {H.~S.}\ \bibnamefont {Reall}}, \
  and\ \bibinfo {author} {\bibfnamefont {J.~E.}\ \bibnamefont {Santos}},\
  }\href {\doibase 10.1088/1361-6382/aafcf2} {\bibfield  {journal} {\bibinfo
  {journal} {Class. Quant. Grav.}\ }\textbf {\bibinfo {volume} {36}},\ \bibinfo
  {pages} {045005} (\bibinfo {year} {2019})},\ \Eprint
  {http://arxiv.org/abs/1808.04832} {arXiv:1808.04832 [gr-qc]} \BibitemShut
  {NoStop}%
\bibitem [{\citenamefont {Hod}(2018)}]{Hod:2018}%
  \BibitemOpen
  \bibfield  {author} {\bibinfo {author} {\bibfnamefont {S.}~\bibnamefont
  {Hod}},\ }\href {\doibase 10.1016/j.physletb.2018.09.039} {\bibfield
  {journal} {\bibinfo  {journal} {Phys. Lett. B}\ }\textbf {\bibinfo {volume}
  {786}},\ \bibinfo {pages} {217} (\bibinfo {year} {2018})},\ \bibinfo {note}
  {[Erratum: Phys.Lett.B 796, 256 (2019)]},\ \Eprint
  {http://arxiv.org/abs/1808.04077} {arXiv:1808.04077 [gr-qc]} \BibitemShut
  {NoStop}%
\bibitem [{\citenamefont {Cardoso}\ \emph
  {et~al.}(2018{\natexlab{b}})\citenamefont {Cardoso}, \citenamefont {Costa},
  \citenamefont {Destounis}, \citenamefont {Hintz},\ and\ \citenamefont
  {Jansen}}]{Cardoso:2018}%
  \BibitemOpen
  \bibfield  {author} {\bibinfo {author} {\bibfnamefont {V.}~\bibnamefont
  {Cardoso}}, \bibinfo {author} {\bibfnamefont {J.~L.}\ \bibnamefont {Costa}},
  \bibinfo {author} {\bibfnamefont {K.}~\bibnamefont {Destounis}}, \bibinfo
  {author} {\bibfnamefont {P.}~\bibnamefont {Hintz}}, \ and\ \bibinfo {author}
  {\bibfnamefont {A.}~\bibnamefont {Jansen}},\ }\href {\doibase
  10.1103/PhysRevD.98.104007} {\bibfield  {journal} {\bibinfo  {journal} {Phys.
  Rev. D}\ }\textbf {\bibinfo {volume} {98}},\ \bibinfo {pages} {104007}
  (\bibinfo {year} {2018}{\natexlab{b}})},\ \Eprint
  {http://arxiv.org/abs/1808.03631} {arXiv:1808.03631 [gr-qc]} \BibitemShut
  {NoStop}%
\bibitem [{\citenamefont {Klein}\ \emph {et~al.}(2021)\citenamefont {Klein},
  \citenamefont {Hollands},\ and\ \citenamefont {Zahn}}]{supplement}%
  \BibitemOpen
  \bibfield  {author} {\bibinfo {author} {\bibfnamefont {C.}~\bibnamefont
  {Klein}}, \bibinfo {author} {\bibfnamefont {S.}~\bibnamefont {Hollands}}, \
  and\ \bibinfo {author} {\bibfnamefont {J.}~\bibnamefont {Zahn}},\ }\href
  {https://journals.aps.org/prl/abstract/10.1103/PhysRevLett.127.231301#supplemental}
  {\enquote {\bibinfo {title} {Quantum (dis)charge of black hole
  interiors-supplementary material},}\ } (\bibinfo {year} {2021})\BibitemShut
  {NoStop}%
\bibitem [{\citenamefont {G\'erard}\ and\ \citenamefont
  {Wrochna}(2016)}]{Gerard:2014}%
  \BibitemOpen
  \bibfield  {author} {\bibinfo {author} {\bibfnamefont {C.}~\bibnamefont
  {G\'erard}}\ and\ \bibinfo {author} {\bibfnamefont {M.}~\bibnamefont
  {Wrochna}},\ }\href {\doibase 10.2140/apde.2016.9.111} {\bibfield  {journal}
  {\bibinfo  {journal} {Anal. Part. Diff. Eq.}\ }\textbf {\bibinfo {volume}
  {9}},\ \bibinfo {pages} {111} (\bibinfo {year} {2016})},\ \Eprint
  {http://arxiv.org/abs/1409.6691} {arXiv:1409.6691 [math-ph]} \BibitemShut
  {NoStop}%
\bibitem [{\citenamefont {Verch}(1994)}]{Verch:1992}%
  \BibitemOpen
  \bibfield  {author} {\bibinfo {author} {\bibfnamefont {R.}~\bibnamefont
  {Verch}},\ }\href {\doibase 10.1007/BF02173427} {\bibfield  {journal}
  {\bibinfo  {journal} {Commun. Math. Phys.}\ }\textbf {\bibinfo {volume}
  {160}},\ \bibinfo {pages} {507} (\bibinfo {year} {1994})}\BibitemShut
  {NoStop}%
\bibitem [{\citenamefont {Hawking}(1975)}]{Hawking:1974}%
  \BibitemOpen
  \bibfield  {author} {\bibinfo {author} {\bibfnamefont {S.}~\bibnamefont
  {Hawking}},\ }\href {\doibase 10.1007/BF02345020} {\bibfield  {journal}
  {\bibinfo  {journal} {Commun. Math. Phys.}\ }\textbf {\bibinfo {volume}
  {43}},\ \bibinfo {pages} {199} (\bibinfo {year} {1975})},\ \bibinfo {note}
  {[Erratum: Commun.Math.Phys. 46, 206 (1976)]}\BibitemShut {NoStop}%
\bibitem [{\citenamefont {Gibbons}(1975)}]{Gibbons:1975}%
  \BibitemOpen
  \bibfield  {author} {\bibinfo {author} {\bibfnamefont {G.}~\bibnamefont
  {Gibbons}},\ }\href {\doibase 10.1007/BF01609829} {\bibfield  {journal}
  {\bibinfo  {journal} {Commun. Math. Phys.}\ }\textbf {\bibinfo {volume}
  {44}},\ \bibinfo {pages} {245} (\bibinfo {year} {1975})}\BibitemShut
  {NoStop}%
\bibitem [{\citenamefont {Unruh}(1976)}]{Unruh:1976}%
  \BibitemOpen
  \bibfield  {author} {\bibinfo {author} {\bibfnamefont {W.}~\bibnamefont
  {Unruh}},\ }\href {\doibase 10.1103/PhysRevD.14.870} {\bibfield  {journal}
  {\bibinfo  {journal} {Phys. Rev. D}\ }\textbf {\bibinfo {volume} {14}},\
  \bibinfo {pages} {870} (\bibinfo {year} {1976})}\BibitemShut {NoStop}%
\bibitem [{\citenamefont {Schwinger}(1951)}]{Schwinger:1951}%
  \BibitemOpen
  \bibfield  {author} {\bibinfo {author} {\bibfnamefont {J.~S.}\ \bibnamefont
  {Schwinger}},\ }\href {\doibase 10.1103/PhysRev.82.664} {\bibfield  {journal}
  {\bibinfo  {journal} {Phys. Rev.}\ }\textbf {\bibinfo {volume} {82}},\
  \bibinfo {pages} {664} (\bibinfo {year} {1951})}\BibitemShut {NoStop}%
\bibitem [{\citenamefont {Herman}\ and\ \citenamefont
  {Hiscock}(1994)}]{Herman:1994}%
  \BibitemOpen
  \bibfield  {author} {\bibinfo {author} {\bibfnamefont {R.}~\bibnamefont
  {Herman}}\ and\ \bibinfo {author} {\bibfnamefont {W.~A.}\ \bibnamefont
  {Hiscock}},\ }\href {\doibase 10.1103/PhysRevD.49.3946} {\bibfield  {journal}
  {\bibinfo  {journal} {Phys. Rev. D}\ }\textbf {\bibinfo {volume} {49}},\
  \bibinfo {pages} {3946} (\bibinfo {year} {1994})}\BibitemShut {NoStop}%
\bibitem [{\citenamefont {Sorkin}\ and\ \citenamefont
  {Piran}(2001)}]{Sorkin:2000}%
  \BibitemOpen
  \bibfield  {author} {\bibinfo {author} {\bibfnamefont {E.}~\bibnamefont
  {Sorkin}}\ and\ \bibinfo {author} {\bibfnamefont {T.}~\bibnamefont {Piran}},\
  }\href {\doibase 10.1103/PhysRevD.63.084006} {\bibfield  {journal} {\bibinfo
  {journal} {Phys. Rev. D}\ }\textbf {\bibinfo {volume} {63}},\ \bibinfo
  {pages} {084006} (\bibinfo {year} {2001})},\ \Eprint
  {http://arxiv.org/abs/gr-qc/0009095} {arXiv:gr-qc/0009095} \BibitemShut
  {NoStop}%
\bibitem [{\citenamefont {Poisson}\ and\ \citenamefont
  {Israel}(1989)}]{Poisson:1989}%
  \BibitemOpen
  \bibfield  {author} {\bibinfo {author} {\bibfnamefont {E.}~\bibnamefont
  {Poisson}}\ and\ \bibinfo {author} {\bibfnamefont {W.}~\bibnamefont
  {Israel}},\ }\href {\doibase 10.1103/PhysRevLett.63.1663} {\bibfield
  {journal} {\bibinfo  {journal} {Phys. Rev. Lett.}\ }\textbf {\bibinfo
  {volume} {63}},\ \bibinfo {pages} {1663} (\bibinfo {year}
  {1989})}\BibitemShut {NoStop}%
\bibitem [{\citenamefont {Hollands}\ and\ \citenamefont
  {Wald}(2015)}]{Hollands:2014}%
  \BibitemOpen
  \bibfield  {author} {\bibinfo {author} {\bibfnamefont {S.}~\bibnamefont
  {Hollands}}\ and\ \bibinfo {author} {\bibfnamefont {R.~M.}\ \bibnamefont
  {Wald}},\ }\href {\doibase 10.1016/j.physrep.2015.02.001} {\bibfield
  {journal} {\bibinfo  {journal} {Phys. Rept.}\ }\textbf {\bibinfo {volume}
  {574}},\ \bibinfo {pages} {1} (\bibinfo {year} {2015})},\ \Eprint
  {http://arxiv.org/abs/1401.2026} {arXiv:1401.2026 [gr-qc]} \BibitemShut
  {NoStop}%
\bibitem [{\citenamefont {Lanir}\ \emph {et~al.}(2018)\citenamefont {Lanir},
  \citenamefont {Levi}, \citenamefont {Ori},\ and\ \citenamefont
  {Sela}}]{Lanir:2017}%
  \BibitemOpen
  \bibfield  {author} {\bibinfo {author} {\bibfnamefont {A.}~\bibnamefont
  {Lanir}}, \bibinfo {author} {\bibfnamefont {A.}~\bibnamefont {Levi}},
  \bibinfo {author} {\bibfnamefont {A.}~\bibnamefont {Ori}}, \ and\ \bibinfo
  {author} {\bibfnamefont {O.}~\bibnamefont {Sela}},\ }\href {\doibase
  10.1103/PhysRevD.97.024033} {\bibfield  {journal} {\bibinfo  {journal} {Phys.
  Rev. D}\ }\textbf {\bibinfo {volume} {97}},\ \bibinfo {pages} {024033}
  (\bibinfo {year} {2018})},\ \Eprint {http://arxiv.org/abs/1710.07267}
  {arXiv:1710.07267 [gr-qc]} \BibitemShut {NoStop}%
\bibitem [{\citenamefont {Sela}(2018)}]{Sela:2018}%
  \BibitemOpen
  \bibfield  {author} {\bibinfo {author} {\bibfnamefont {O.}~\bibnamefont
  {Sela}},\ }\href {\doibase 10.1103/PhysRevD.98.024025} {\bibfield  {journal}
  {\bibinfo  {journal} {Phys. Rev. D}\ }\textbf {\bibinfo {volume} {98}},\
  \bibinfo {pages} {024025} (\bibinfo {year} {2018})},\ \Eprint
  {http://arxiv.org/abs/1803.06747} {arXiv:1803.06747 [gr-qc]} \BibitemShut
  {NoStop}%
\bibitem [{\citenamefont {Klein}\ and\ \citenamefont {Zahn}(2021)}]{PRDpaper}%
  \BibitemOpen
  \bibfield  {author} {\bibinfo {author} {\bibfnamefont {C.}~\bibnamefont
  {Klein}}\ and\ \bibinfo {author} {\bibfnamefont {J.}~\bibnamefont {Zahn}},\
  }\href {\doibase 10.1103/PhysRevD.104.025009} {\bibfield  {journal} {\bibinfo
   {journal} {Phys. Rev. D}\ }\textbf {\bibinfo {volume} {104}},\ \bibinfo
  {pages} {025009} (\bibinfo {year} {2021})},\ \Eprint
  {http://arxiv.org/abs/2104.06005} {arXiv:2104.06005 [gr-qc]} \BibitemShut
  {NoStop}%
\bibitem [{\citenamefont {Suzuki}\ \emph {et~al.}(1999)\citenamefont {Suzuki},
  \citenamefont {Takasugi},\ and\ \citenamefont {Umetsu}}]{Suzuki:1999}%
  \BibitemOpen
  \bibfield  {author} {\bibinfo {author} {\bibfnamefont {H.}~\bibnamefont
  {Suzuki}}, \bibinfo {author} {\bibfnamefont {E.}~\bibnamefont {Takasugi}}, \
  and\ \bibinfo {author} {\bibfnamefont {H.}~\bibnamefont {Umetsu}},\ }\href
  {\doibase 10.1143/PTP.102.253} {\bibfield  {journal} {\bibinfo  {journal}
  {Prog.\ Theor.\ Phys.}\ }\textbf {\bibinfo {volume} {102}},\ \bibinfo {pages}
  {253} (\bibinfo {year} {1999})},\ \Eprint
  {http://arxiv.org/abs/gr-qc/9905040} {arXiv:gr-qc/9905040} \BibitemShut
  {NoStop}%
\bibitem [{\citenamefont {Hatsuda}(2020)}]{Hatsuda:2020}%
  \BibitemOpen
  \bibfield  {author} {\bibinfo {author} {\bibfnamefont {Y.}~\bibnamefont
  {Hatsuda}},\ }\href {\doibase 10.1088/1361-6382/abc82e} {\bibfield  {journal}
  {\bibinfo  {journal} {Class. Quant. Grav.}\ }\textbf {\bibinfo {volume}
  {38}},\ \bibinfo {pages} {025015} (\bibinfo {year} {2020})},\ \Eprint
  {http://arxiv.org/abs/2006.08957} {arXiv:2006.08957 [gr-qc]} \BibitemShut
  {NoStop}%
\end{thebibliography}%
\end{document}